\documentclass[10pt,journal,cspaper,compsoc]{IEEEtran}
\usepackage{graphicx}
\usepackage{amssymb}
\usepackage{url} 
\usepackage[normalem]{ulem}
\usepackage[dvips]{color}
\usepackage{booktabs} 
\usepackage{psfrag} 
\usepackage{stfloats} 
\ifCLASSOPTIONcompsoc
	\usepackage[nocompress]{cite}
\else
	\usepackage{cite}
\fi
\begin{document}
\title{GReTA~~--~~a novel Global and Recursive\\Tracking Algorithm in three dimensions}
\author{Alessandro~Attanasi,~Andrea~Cavagna,~Lorenzo~Del~Castello,~Irene~Giardina,~Asja~Jeli\'{c},\\
		Stefania~Melillo,~Leonardo~Parisi,~Fabio~Pellacini,~Edward~Shen,~Edmondo~Silvestri,~Massimiliano~Viale
\IEEEcompsocitemizethanks{\IEEEcompsocthanksitem The authors are with the Istituto Sistemi Complessi, Consiglio Nazionale delle Ricerche, UOS Sapienza, 00185 Rome, Italy.\protect\\E-mail: see http://www.cobbs.it%
\IEEEcompsocthanksitem A.~Attanasi, A.~Cavagna, L.~Del~Castello, I.~Giardina, A.~Jeli\'{c}, S.~Melillo, and M.~Viale are with the Dipartimento di Fisica, Universit\`{a} Sapienza, 00185 Rome, Italy.%
\IEEEcompsocthanksitem L.~Parisi and F.~Pellacini are with Dipartimento di Informatica, Universit\`{a} Sapienza, 00198 Rome, Italy.%
\IEEEcompsocthanksitem E.~Shen is with Bublcam Technology Inc., Toronto, Canada.%
\IEEEcompsocthanksitem E.~Silvestri is with the Dipartimento di Matematica e Fisica, Universit\`{a} Roma Tre, 00146 Rome, Italy.}%
\thanks{}}
%
%
%
\IEEEcompsoctitleabstractindextext{
	\begin{abstract}
		Tracking multiple moving targets allows quantitative measure of the dynamic behavior in systems as diverse as animal groups in biology, turbulence in fluid dynamics and crowd and traffic control. In three dimensions, tracking several targets becomes increasingly hard since optical occlusions are very likely, i.e. two featureless targets frequently overlap for several frames. Occlusions are particularly frequent in biological groups such as bird flocks, fish schools, and insect swarms, a fact that has severely limited collective animal behavior field studies in the past. This paper presents a 3D tracking method that is robust in the case of severe occlusions.
		To ensure robustness, we adopt a global optimization approach that works on all objects and frames at once. To achieve practicality and scalability, we employ a divide and conquer formulation, thanks to which the computational complexity of the problem is reduced by orders of magnitude. We tested our algorithm with synthetic data, with experimental data of bird flocks and insect swarms and with public benchmark datasets, and show that our system yields high quality trajectories for hundreds of moving targets with severe overlap. The results obtained on very heterogeneous data show the potential applicability of our method to the most diverse experimental situations.
	\end{abstract}
}
\maketitle
\IEEEdisplaynotcompsoctitleabstractindextext
\IEEEpeerreviewmaketitle
\ifCLASSOPTIONcompsoc
	\noindent\raisebox{2\baselineskip}[0pt][0pt]%
	{\parbox{\columnwidth}{\section{Introduction}\label{sec:introduction}%
	\global\everypar=\everypar}}%
	\vspace{-1\baselineskip}\vspace{-\parskip}\par
\else
	\section{Introduction}\label{sec:introduction}\par
\fi
		\IEEEPARstart{I}{n} recent years there has been a growing interest in studying the motion of large groups of objects, both in two and in three dimensions: animals, humans, automotive vehicles, cells and microorganisms in field or laboratory experiments, as well as tracer particles in turbulent fluids flows~\cite{adrian1991arfm,luthi2005jfm,ouellette2006eif,cullen1965ab,dahmen1984prsb,pomeroy1992auk,malik1993eif,doh2000eif,willneff2002istpdrm,betke2009iccv,zou2009Aiccv,wu2009Biccv,wu2011oe,wu2011cvpr,liu2012eccv,ardekani2012jrs}. This kind of studies requires tracking, the automated process of following in space and time individual objects using visual information from single- or multi-camera video sequences.
		\begin{figure*}
			\includegraphics[width=1.99\columnwidth]{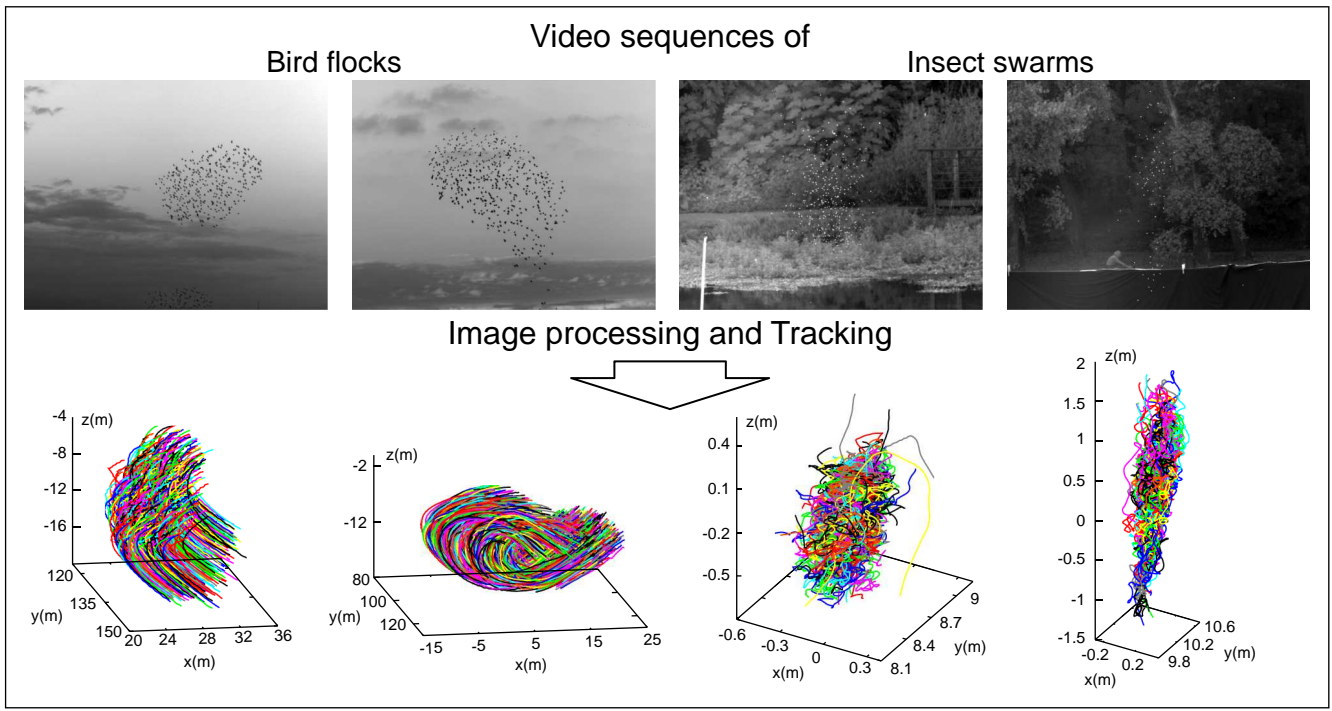}
			\caption{\small Input and output data of our tracking algorithm. Top row: examples of original images extracted from the video sequences taken during four field experiments of flocking birds and swarming insects (from left to right, experimental events $E3$, $E6$, $E14$, and $E15$, respectively. See Section~\ref{sec:expdata} and Table~\ref{tab:exp} therein). The original images are slightly cropped and enhanced for the sake of readability. Bottom row: the 3D reconstruction of the full trajectories for each experimental event.}
			\label{fig1}
		\end{figure*}
		Sometimes experimental data contains information on the object's features, so that for example color- or pattern-matching strategies can be exploited to simplify the problem. This, however, is not our case: we focus here on the three-dimensional tracking problem using stereometric information only. Examples of input and output data of our tracking algorithm are shown in Fig.~\ref{fig1}.

		There are two main reasons why tracking is hard. First, when the average inter-object distance in the images is small compared to the average displacement of the objects between two consecutive frames, ambiguities arise when identifying individual objects in time. This is easily solved using cameras with a sufficiently high temporal resolution. 

		A second and far more serious difficulty arises when the average inter-objects distance in the images is small compared to their optical size, making optical occlusions highly likely. Each time an ambiguity due to an occlusion occurs, there is a high probability that the tracked trajectories of the objects involved are interrupted.
		
		These interruptions are a minor problem when estimating velocity fields, but the situation becomes more problematic when we use the velocities to infer the inter-individual interactions within a group of animals. Several interruptions at any given time frame are equivalent to missing some of the individuals, which potentially biases the inferred interaction. The problem is even more serious when we measure observables that depend on the \textit{entire} individual trajectories such as diffusion properties~\cite{cavagna2013prslb} or the kinematics of turning~\cite{attanasi2014information} in collective animal behavior. Even in turbulence studies, the lack of complete trajectories can introduce serious statistical biases on some physical observables~\cite{biferale2008pof}.
		
		In fact, interruptions are the best case scenario when we have many optical occlusions. The worst case is the introduction of non-existent trajectories that mix the identities of two different objects, especially problematic in physical and biological analysis.
		\subsection{Literature survey}\label{sec:literature}
			Tracking algorithms differ according to how they exploit the information available in the images.
			In the last thirty years, several algorithms have been developed in the field of fluid dynamics~\cite{adrian1991arfm,luthi2005jfm,ouellette2006eif}. In particular, the algorithm by Ouellette~\textit{et~al.}~\cite{ouellette2006eif}, in case of ambiguities, optimizes the solution for all tracked particles locally in time. In the field of collective animal behavior, different algorithms have been developed to reconstruct the 3D-positions of individual animals in groups~\cite{cullen1965ab,dahmen1984prsb,pomeroy1992auk}.
			The technical shortcomings and limited results of these initial studies have been the catalyst for the successive empirical investigations on collective animal phenomena, and lead to the development of tracking algorithms tested on various animals as fruit flies~\cite{straw2010multi,liu2012eccv,zou2009Aiccv}, mosquitoes~\cite{butail20113d, butail2012reconstructing}, bees~\cite{veeraraghavan2006motion}, bats~\cite{betke2009iccv,wu2011cvpr,wu2009tracking}, and fish~\cite{butail20103d}.

			A recent significant breakthrough in the field is represented by the Multiple Hypothesis Tracking (MHT) approach~\cite{reid1979algorithm}, which finds objects correspondences across multiple views through an NP-hard multidimensional assignment problem.
			MHT methods based on global optimization over the space of the tracked objects and/or time have been implemented with greedy approaches. Betke~\textit{et~al.}~\cite{betke2009iccv} developed an algorithm based on a multi-dimensional assignment problem solved with a greedy approximation. Zou~\textit{et~al.}~\cite{zou2009Aiccv} implemented a tracking algorithm which uses a global correspondence selection scheme, and applies Gibbs sampling locally in time to reduce the complexity of the algorithm. H.S. Wu~\textit{et~al.}~\cite{wu2009Biccv,wu2011oe} implemented a different algorithm based on three linear assignment problems, making use of ghost objects to partially solve the problem of short-term optical occlusions.
			Liu ~\textit{et~al.}~\cite{liu2012eccv} proposed a very efficient algorithm in the framework of particle filter able to deploy weak visual information to distinguish the identities of the tracked objects. More interesting is the approach proposed by Z. Wu~\textit{et~al.}~\cite{wu2011cvpr}, who recognized the importance of a global optimization over the full temporal sequence and over all the tracked objects, posing the problem in the form of a weighted set-cover.

			Our efforts focus on 3D tracking of large groups of featureless objects for long temporal sequences, and the long-term optical occlusions typical of our experimental data need to be addressed in a different way. Therefore we aim at the globally optimal solution of the problem, and not at efficient and fast ways to approximate it with greedy approaches. The bottleneck of this strategy is the computational complexity which grows exponentially with the duration of the acquisition.
		\subsection{Our tracking approach}\label{sec:ourtrackingapproach}
			We propose here a novel Global and Recursive Tracking Algorithm (GReTA), an approach which dramatically reduces the computational complexity of the global optimization problem thanks to a recursive divide and conquer strategy. Within this new framework, we can optimize the solution globally over longer temporal sequences. In order to preserve the global scope, we introduce a way to extend the temporal horizon over which ambiguous choices are made within the divide and conquer scheme.

			Thanks to this method, we are able to resolve optical occlusions lasting up to dozens of consecutive frames, and therefore to distinguish the identities of the tracked objects without creating interruptions, even when the optical density in the images is very large.
			The reconstructed trajectories have negligible fragmentation even in the presence of large optical density and frequent occlusions.
			
			We validate our approach using synthetic data as ground-truth, and we test its potential by applying it to original experimental field data of flocking birds and swarming insects.

			The rest of the paper is divided into 6 sections. Section 2 explains the algorithm. In Section 3 we analyze the problem complexity. Sections 4 and 5 report the validation of our algorithm with synthetic data and experimental field data. In Section 6 we present a comparison with prior works, and the conclusions in Section 7.

	\section{Methods}\label{sec:methods}
		The main idea of our method is that when an occlusion occurs, we assign multiple temporal links and we use these links to create all possible paths running through the occlusion. Many of these paths will be non-physical, but certainly the paths corresponding to the real objects will also be there. Then, the information from all cameras is assembled and the selection of the physically meaningful paths, namely those that optimize multi-camera coherence, is performed globally in space (over the tracked objects) and in time (over the temporal sequence), making use of a recursive divide and conquer scheme.
		\subsection{The basic steps of a tracking system}\label{sec:trackingtasks}
			\begin{figure*}
				\includegraphics[width=1.99\columnwidth]{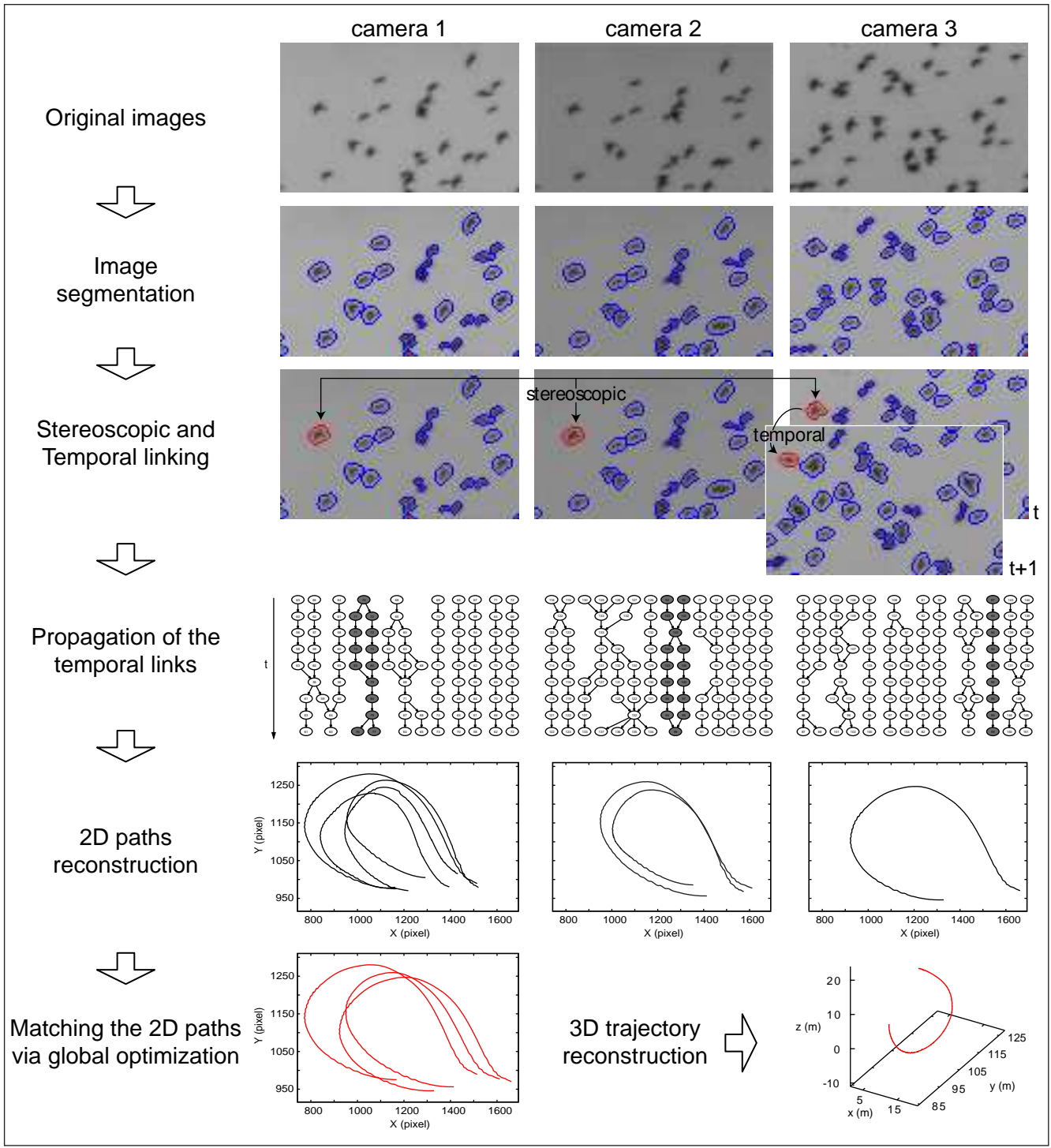}
				\caption{\small Scheme illustrating the main steps of the full tracking system (image processing and tracking algorithms). A small crop of the original images extracted from the video sequences of three synchronized cameras are shown on the first row. The segmented images are shown on the second row using blue color for the object borders, and red for their centers of mass. On the third row, we show one example of a (trifocal) stereoscopic link connecting the views of the same object in the three cameras, and one example of temporal link connecting the same object between subsequent frames in each camera sequence. In the fourth row, we show a crop of the temporal graphs for each camera view, which represents a useful visualization of the full set of temporal links assigned for each camera.
				The figures show only a small crop representing a few objects for $9$ time frames, and we indicated with grey color a cluster of paths stereoscopically linked across the three views. The fifth row illustrates the 2D paths reconstructed in the image space of each camera, obtained by simple propagation of the temporal links. The algorithm outputs all possible 2D paths at this step. Global optimization is used to match the correct 2D paths between the camera views, as shown on the sixth and last row, from which we finally obtain the 3D the trajectory using standard stereoscopic geometry.}
				\label{fig2}
			\end{figure*}
			The goal is to track individual objects in time while reconstructing their positions in 3D space. We use stereoscopic video-sequences of the target objects acquired via a synchronized and calibrated three-camera system. The data gathering procedures we used in our experiments are described in Appendix~A.
			\\\\
			\textbf{Image segmentation.} The first step of a tracking algorithm is the detection of the objects in the images, done by image segmentation, see Fig.~\ref{fig2}, first and second rows. Several approaches may be used to perform the segmentation, and the choice strongly depends on the type of objects. Our approach to image segmentation is not an essential part of the tracking system we propose, so we leave its description to Appendix~B.
			\\\\
			\textbf{Stereoscopic linking.} The second step is to compute the stereoscopic linking of the detected objects, which consists of matching the individual objects across the images acquired by different cameras at the same time, see Fig.~\ref{fig2}, third row. We assign multiple stereoscopic links between the object images as seen by three cameras using standard trifocal geometry~\cite{hartley2003book}. The details of the linking method do not matter, and we describe the exact procedure in Appendix~C.
			\\\\
			\textbf{Temporal linking.} The third step of our algorithm is to assign multiple temporal links for each object, which consists of matching individual objects from one frame to the next one, as shown on the same third row of Fig.~\ref{fig2}. We use different prediction strategies according to the specific data we process. The precise details of the linking methods are not essential, and the exact procedures are described in Appendix~C.
			\\\\
			\textbf{Tracking.} Recent global optimization approaches, as the one we propose here, rely on the assumption that objects may be linked to several other objects (multi-linking instead of one-to-one linking), and the global optimization is performed over the space of these links to select the matches corresponding to real 3D trajectories. In the following sections, we explain the method and we present the formalisms of our tracking approach.
		\subsection{Multi-path branching}\label{sec:multipath}
			\begin{figure*}
				\includegraphics[width=0.99\textwidth]{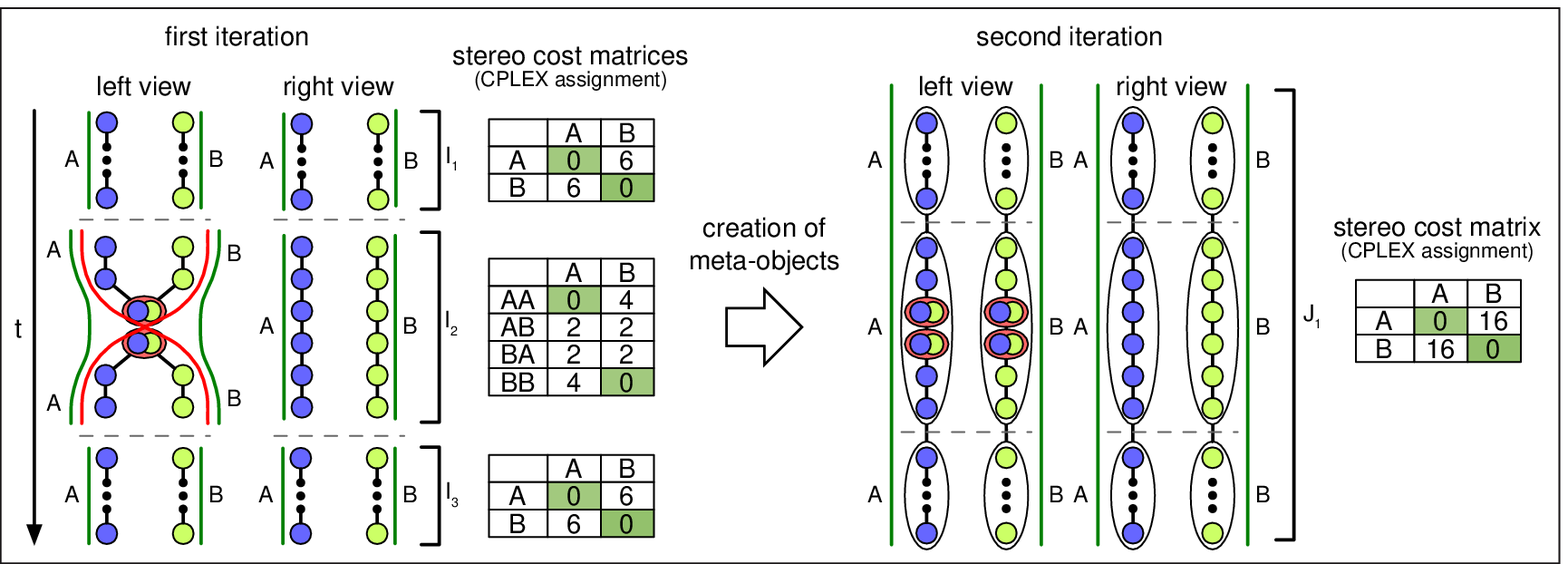}
				\caption{\small Scheme illustrating a real case of confined data. Two objects $A$ and $B$ occlude each other for two frames in the left view, while they always appear as separate objects in the right view. During the first iteration of the recursive divide and conquer approach, the event is divided into three intervals, $I_1$, $I_2$, and $I_3$. In the first and in the third intervals, there are no tracking ambiguities. The propagation of the temporal links results only in the two correct 2D paths, $A$ and $B$, in each camera. We define the cost of a pair of 2D paths as the sum of the costs of the links between them, i.e. the number of missing stereoscopic links. The global optimization selects the correct matches, $(A,A)$ and $(B,B)$.
				In the second interval, there are still not ambiguities in the right view: the only 2D paths, $A$ and $B$, are correct. Instead, in the left view, we propagate the temporal links and we create $4$ 2D paths; the two correct ones (green lines) $AA$ and $BB$, and the two wrong ones (red lines) $AB$ and $BA$. Two of the possible set-cover solutions $\Gamma$ are: the correct one $G\equiv\{(AA,A), (BB,B)\}$ with a cost equal to $0$, and the wrong one $\{(AB,A), (BA,B)\}$ with a cost equal to $4$. Here, the global optimization is essential to select the correct solution.
				All the matched 2D paths are then analyzed at the second iteration as meta-objects. At this iteration, there are no occlusions. Propagating the temporal links, we create two 2D paths in each camera view, and the tracking problem is correctly solved for the entire duration of the event.}
				\label{fig3}
			\end{figure*}
			Let us consider the example shown in Fig.~\ref{fig3}, which illustrates a partial temporal sequence of two objects $A$ and $B$ as seen by two cameras. The two objects overlap in the image of the left camera for three frames. Most prior work would assign only one temporal link to each detected object, therefore the points of occlusion belong to only one uninterrupted trajectory, the second recovered trajectory being broken. This results in a fragmented trajectory. Furthermore, assigning temporal links using information that is purely local has the drawback that the identities of occluding objects might be lost.
			\\\\
			To tackle this concerns, we use a path branching approach with global optimization. For the example in Fig.~\ref{fig3}, we assign multiple temporal links and create all possible paths running through the occlusion in the left camera view. In this case, there are four paths, of which two are real ($AA$ and $BB$) and two have hybrid object identities ($AB$ and $BA$). In order to build the set of all possible paths through the segmented objects, the temporal links are propagated for each camera view to build the temporal graph of each camera. 
			We then have to solve the problem of how to select the correct paths in the 2D graph of each camera and match them across cameras. The advantage of our approach is that at an early stage each object can have more than one path, which is what is needed to handle occlusions.
		\subsection{Global optimization}\label{sec:globalopt}
			The selection of the correct matches between the 2D paths across cameras is the core of the tracking problem. We create all the possible 2D paths propagating the temporal links in the image space of each camera, while we choose how to match them using stereoscopic links. The assumption here is that the 2D paths representing the same 3D object are strongly linked stereoscopically. On the contrary, 2D paths corresponding to different 3D objects are loosely linked stereoscopically. We define a measure of the stereoscopic quality of each match, \textit{i.e.} a cost function, and we use a global optimization approach to retrieve the set of the correct matches. This is an NP-hard multidimensional assignment problem, and we solve it using Integer Linear Programming (ILP) techniques in order to find the globally optimal solution~\cite{fisher2004lagrangian}.
			\\\\
			\textbf{Definition of trajectory.} Consider a system of three cameras, and denote a trajectory $\gamma$ as a triplet of matched 2D paths, $\gamma=(\gamma_1,\gamma_2,\gamma_3)$. Each 2D path $\gamma_i$ represents a temporal sequence of 2D objects detected in the images of the $i$-th camera, and connected by temporal links. Moreover the triplet $(\gamma_1, \gamma_2, \gamma_3)$ is stereoscopically linked for at least one frame. Let $\Gamma$ be the set of all the possible trajectories. The goal is to find the correct subset of trajectories $G\subseteq\Gamma$, see Fig.~\ref{fig3}.
			\\\\
			\textbf{Cost function.} We evaluate the quality of each subset $\widehat{\Gamma}\subseteq\Gamma$ by defining a cost function $C(\widehat{\Gamma})$. Let $C(\widehat{\Gamma})=\sum_{\gamma\in\widehat{\Gamma}}c(\gamma)$, where $c(\gamma)$ is a cost associated to the trajectory $\gamma$ and based on the stereoscopic coherence (the higher the quality, the lower the cost), see Fig.~\ref{fig3}.

			Let us formally define the cost function. Considering a three-camera system, for each $\gamma\in\Gamma$ and at each instant of time, the cost function $c(\gamma(t))=c(\gamma_1(t),\gamma_2(t),\gamma_3(t))$ is defined as the trifocal distance~\cite{hartley2003book} in the case of matched triplets, and as the epipolar distance~\cite{hartley2003book} in the case of pairs (corresponding to miss-detection in one camera). Whenever the cost exceeds a threshold value $c_{max}$, or in the case of absence of a stereoscopic link, we set $c=c_{max}$. The cost of a trajectory $\gamma$ is then defined as the temporal average:
			\begin{equation}
				c(\gamma)=\frac{\sum_{t\in T_{\gamma}}c(\gamma(t))}{|T_{\gamma}|},
				\label{eqn:costfunction}
			\end{equation}
			where $T_\gamma$ is the set of time frames, $t$, for which $c(\gamma(t))$ is defined.
			\subsubsection{Formalization of the tracking problem}\label{sec:formalism}
				Let us distinguish between two different types of input data.
				\begin{description}
					\item[{\it Confined data}:]\hspace{30pt} the objects are in the common field-of-view of the camera system at least for a short temporal sequence. Each segmented object belongs to at least one trajectory, and the solution of the tracking problem is a cover for the set of all the objects.
					\item[{\it Non-confined data:}]\hspace{48pt} one or more objects never appear in the common field-of-view of the camera system, but they are seen by one camera only. Therefore they are far from the objects of interest in three dimensions, and they should not be matched as they do not belong to any trajectory. A typical example is represented by pollen particles passing in front of one camera only, and appearing as large blurred objects. The problem becomes more complex, and the covering condition needs to be relaxed to exclude these objects.
				\end{description}
				\noindent\textbf{Confined data, joint weighted set-cover.} When applied to confined data, the global optimization approach is equivalent to a joint weighted set-cover (as in ~\cite{wu2011cvpr}). The tracking problem can be formulated as:
				\begin{equation}
					c(\Gamma_{opt})=\min_{\lbrace x\rbrace}\sum_{\gamma\in\Gamma}c(\gamma)x_{\gamma}~,
					\label{eqn:minimum}
				\end{equation}
				with the constraint:
				\begin{equation}
					\forall p~,~~\sum_{\gamma\in\Gamma_p}x_{\gamma}\geq1~,
					\label{eqn:constraint}
				\end{equation}
				where $x_\gamma$ is a boolean variable associated to $\gamma$, $p$ is a 2D object in the image space of a camera, and $\Gamma_p$ is the set of all trajectories passing by $p$. The retrieved set $\Gamma_{opt}\equiv\{\gamma~|~x_{\gamma}=1\}$ covers with the best weight the full set of segmented objects.

\begin{figure}[!b]
					\centering
					\includegraphics[width=0.75\columnwidth]{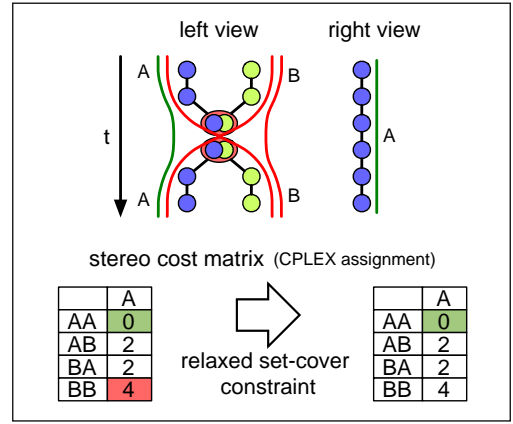}
					\caption{\small Scheme illustrating a real case of non-confined data, i.e. data corrupted by the images of objects in one camera view which do not belong to the group of interest and do not appear in the common field-of-view of all cameras (e.g., insects or birds flying in front of one camera only, or a pollen particle passing in front of a camera lens). A tracked object $A$ and a pollen particle $B$ occlude each other in the left view, while only $A$ is visible in the right view. There are only two possible set-cover solutions, both characterized by the same cost equal to $4$: $\{(AA,A), (BB,A)\}$ and $\{(AB,A), (BA,A)\}$. Both solutions would produce wrong trajectories, and the algorithm would fail to find the correct solution. Relaxing the set-cover constraint, the correct solution is found as the trajectory $(AA,A)$, while the objects belonging only to the 2D path $BB$ in the left view are discarded.}
					\label{fig4}
				\end{figure}

				It can be proven that, under suitable conditions, the global optimization approach finds the correct solution. Indeed, in the case of confined data, when all the correct temporal and stereoscopic links are known and when some particular ambiguities are forbidden (for a formal definition, see Appendix~D), the correct solution of the tracking problem is the only set-cover minimizing the cost defined by Eq.~\ref{eqn:minimum} with the constraint in Eq.~\ref{eqn:constraint}. We refer the reader to Theorem~1 in Appendix~D
				for the exact list of hypotheses holding this statement, together with its proof.
				\\\\
				\textbf{Non-confined data, relaxed joint weighted set-cover.}  In the case of non confined data, not all the segmented objects can be tracked. We need to discard those objects not appearing in all the three cameras, therefore lacking stereoscopic correspondance, because they do not belong to the group of interest. To this aim, we need to relax the covering constraint in Eq.~\ref{eqn:constraint}. Let us introduce for each detected object $p$ a new boolean variable $y_p$. The relaxed joint weighted set-cover problem is then formalized as:
				\begin{equation}
					\min_{\lbrace x,y\rbrace}\left[\sum_{\gamma\in\Gamma}c(\gamma)x_{\gamma}+\frac{\lambda}{T}\sum_{p}\left(1-y_p\right)\right]~,
					\label{eqn:relaxedminimum}
				\end{equation}
				where $T$ is the event duration, and with the constraint:
				\begin{equation}
					\forall p~,~~\sum_{\gamma\in\Gamma_p}x_{\gamma}\geq y_p~.
					\label{eqn:relaxedconstraint}
				\end{equation}
				The contribution of a discarded object, for which $y_p=0$, to the global cost of the solution is $\lambda/T$. Assigning to $\lambda$ a value lower than the highest cost assigned to the stereoscopic links (the threshold value $c_{max}$), we manage to exclude from the solution the objects detected only in one camera view. We experimentally choose $\lambda=0.9c_{max}$. See the example sketched in Fig.~\ref{fig4}.

				In our implementation, the optimization problem is solved using linear programming, for which we use the library in~\cite{cplex1994}.
		\subsection{Recursive divide and conquer}\label{sec:recursion}
			The computational complexity of global optimization problem strongly limits the size of the datasets that can be processed. In order to reduce the complexity, the full temporal sequence can be divided into shorter intervals over which smaller optimization problems can be solved. A well-known method used to join the subtrajectories constructed within limited time windows is the sliding window approach~\cite{wu2009tracking}.
			This approach matches the subtrajectories of the first interval with the ones of the second one, then the ones of the second with the ones of the third interval, repeating the procedure until the full trajectories are recovered. Such approach is very efficient and extremely powerful when applied to sparse data or when the tracked objects can be identified using features like shape, pattern, or color. Its weakness resides in the optimization which is not performed globally in time. For this reason, the identities of the objects can easily be lost whenever treating dense data of featureless objects.

			The GReTA algorithm we propose here is based on a recursive divide and conquer strategy. We divide the acquisition into temporal intervals with length $\tau_1<T$. The optimization described in the previous section is performed in each time interval. Each path -- a temporal sequence of linked 2D objects -- belonging to a selected trajectory becomes then a \textit{meta}-object, which can be linked stereoscopically and in time to other meta-objects (paths). The procedure is then iterated. A new time interval $\tau_2$ is selected, where now $\tau_2$ enumerates the number of intervals of length $\tau_1$, \textit{i.e.} of \textit{meta}-frames. The procedure is applied recursively, until the product of the $\tau$'s of each iteration equals the duration of the entire acquisition, $\tau_1\tau_2\tau_3=T$. Finally, the partial solutions retrieved at each iteration are combined into the solution of the full problem at the last iteration.
			It is possible to prove that, when the conditions that guarantees the uniqueness of the solution (see Section~\ref{sec:formalism} and Appendix~D) are satisfied within each interval of length $\tau_1$, the solution obtained using the recursive approach coincides with the one obtained solving the problem over the entire temporal sequence. The reader is referred to Corollay~1 in Appendix~D.

			Note that the recursive approach offers two key advantages when compared to the classical sliding window one. First, it permits to evaluate the optimization problem for several interval interfaces at once, giving it a more global scope. Second, it allows postponing ambiguous choices at each iteration to later ones, effectively extending the temporal horizon over which these choices are made.

		\subsection{Making the algorithm robust against wrong or missing links}\label{sec:mods}
			Dealing with real data, the sets of temporal and stereoscopic links are affected by noise, which results in missing links and fluctuations of the stereoscopic distances. In some particular situations, the optimization operated within finite intervals of time is not guaranteed to be equivalent to a truly global optimization over the entire temporal sequence. 

			We describe here two modifications of the algorithm that take into account such situations typical of experimental data: a way to postpone ambiguous choices to the next iteration, effectively extending the temporal horizon over which a choice is made, and a way to recognize and re-join fragments of the same trajectory.
			\subsubsection{Postponing ambiguous choices to the following iterations}\label{sec:modsmps}
				The absence of some correct links may lead to one or more trajectories representing unreal objects. These trajectories are characterized by at least one long time gap during which the stereoscopic links are absent.
				
				We detect these cases by using a threshold over the maximum acceptable number of consecutive frames of missing stereoscopic links, and we discard them. We then run the optimization algorithm. Next, we propagate the links over all the discarded 2D objects, creating new 2D paths. These are then passed to the optimization algorithm at the following iteration together with all the matched 2D paths. In this way, we discard any ambiguous choice made locally within any interval at the current iteration, and postpone the decision to the following iteration, effectively extending the temporal window when necessary.

				Such refined algorithm is applied at each iteration. At the last iteration, the trajectories lacking stereoscopic coherence are discarded. New 2D paths are obtained by propagating through the objects which belonged to the discarded trajectories, and they are added to the set of 2D paths. Finally, the set of all the 2D paths is passed to the optimization algorithm running for a second and last time over the full temporal sequence. This time, the trajectories lacking coherence will not be discarded.
			\subsubsection{Joining trajectory fragments}\label{sec:modsscrondo}
				There are two reasons for the algorithm to output correct but fragmented trajectories. First, when a temporal link is missing. Second, when the modification described above breaks a wrong trajectory and reconstructs two correct fragments.
				In both cases, it is possible to re-join fragments of trajectories which are consecutive in time and stereoscopically coherent. We do so after each iteration by connecting fragmented 2D paths in each camera that are stereoscopically connected to the same full 2D path in another camera. Note that in our implementation that is designed for a system of three cameras, we actually match fragments of paths in one camera only with matched pairs of paths in the other two cameras.
		\subsection{Final quality check and 3D reconstruction}\label{sec:qualitycheck}
			In the case of field experiments with freely moving animals, individuals happen to leave the field-of-view of one camera for long times during the recorded events. Furthermore, the noise present in real experimental images often results in errors of the segmentation routine, both miss-detections and over-detections. Because of these reasons, it is not possible to correctly track all the objects for the entire temporal sequence, expecially when the size of the problem is very large and the image data is heavily corrupted with noise.

			To ameliorate these issues, we discard those few trajectories lacking stereoscopic coherence for a considerably long time gap. As shown in the following Section~\ref{sec:synthdata}, this amounts to roughly $4\%$ of the final trajectories for an average-sized dataset ($512$ objects and $500$ frames, see Table~\ref{tab:synth}). We then cut those trajectories, in order to save the fragments which do satisfy the stereoscopic coherence. Such operation results in a minor trajectory fragmentation.
			
			Finally, we are left with a set of matched triplets of 2D paths. Given a triplet of 2D paths, we can reconstruct the corresponding trajectory in 3D by applying standard stereometric formulas~\cite{hartley2003book} to each triplets of synchronous 2D points belonging to the paths, as shown on the last row of Fig.~\ref{fig2}.
	\section{Complexity of the tracking problem}\label{sec:complexity}
		The global optimization requires comparing all the possible solutions and selecting the one that minimizes the cost function. This is a multidimensional assignment, and it is NP-hard. We solve it by finding the globally optimal solution using ILP techniques. We compute the cost of each possible triplet of linked 2D paths $\gamma\in\Gamma$, \textit{i.e.} with at least one stereoscopic link. This implies that the number of variables to handle, $H$, corresponds to the number of possible trajectories, $|\Gamma|$. The parameter $H$ strictly depends on the number $P$ of 2D paths in the graph of each camera, obtained by propagation of the temporal links, and on the stereoscopic links between them. Therefore both $H$ and $P$ depend on the number of objects to be tracked, and they both grow exponentially with the temporal duration of the event. Let us analyze in detail such trend.
		\begin{figure*}[!b]
			\unitlength=1in
			\centering
			\psfrag{N}{\scriptsize{$\mathcal{N}$}}
			\psfrag{H}{\footnotesize{$H$}}
			\psfrag{Hf}{\footnotesize{$H_{full}$}}
			\includegraphics[width=0.99\textwidth]{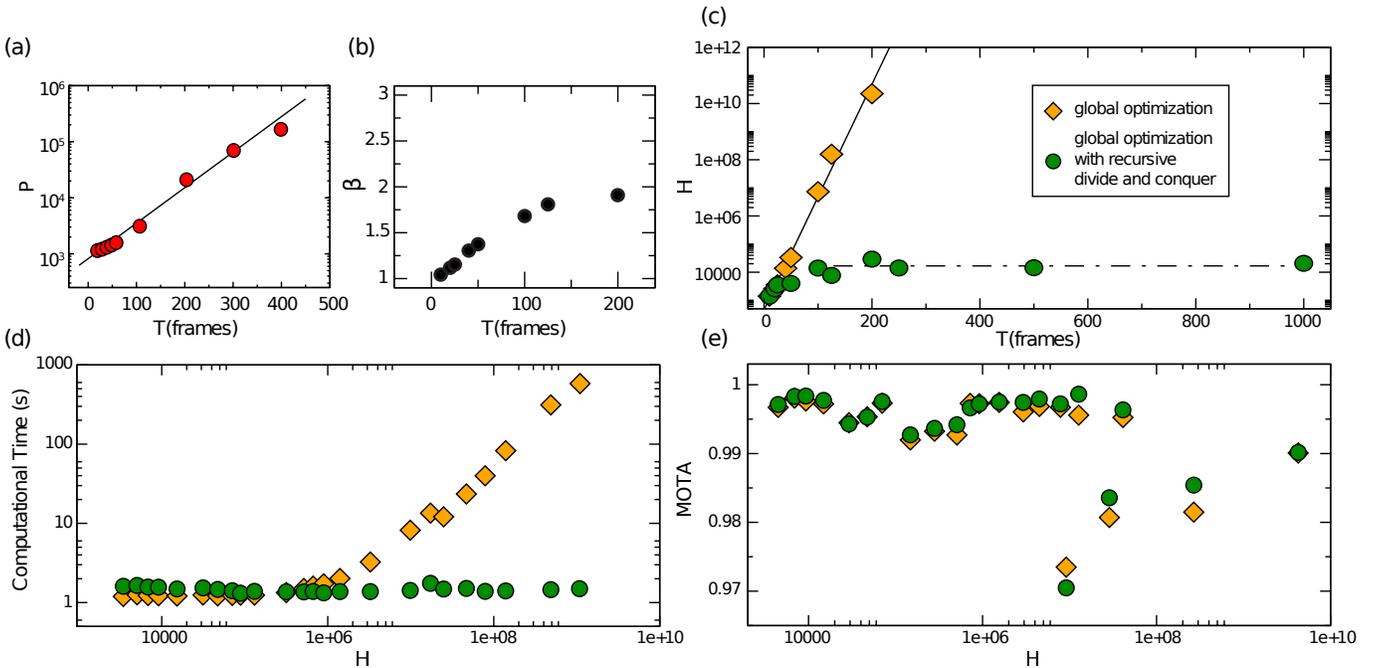}
			\caption{\small The intrinsic complexity of the tracking problem is shown with synthetic data, with and without the recursive divide and conquer scheme is compared.
			In panel ($a$), the number $P$ of possible paths for a synthetic dataset of $1024$ objects is shown as a function of the considered temporal duration $T$.
			In panel($b$), the dependence of $\beta$ on $T$.
			In panel ($c$), the values of $H$ as a function of $T$ are plotted as yellow diamonds; these are compared to the values of $H$ obtained when the recursive divide and conquer strategy is applied (choosing $\tau_1=25$~frames), and plotted as green circles.
			In panel ($d$), the computational time is plotted versus $H$ for several runs with different complexities, with and without recursion as in panel ($c$).
			In panel ($e$), the MOTA (see Sec.~\ref{sec:synthdata}) values obtained on several runs with different complexities are plotted versus $H$, with and without recursion as in panel ($c$).}
			\label{fig5}
		\end{figure*}
		\subsection{Complexity of the temporal graph of each camera space}
			The number of 2D paths, $P$, for a certain camera depends on the temporal length of the acquisition and on the connectivity of the graph on that camera. This dependence can be described by introducing a bifurcation coefficient $\alpha\geq0$, which is an indirect measure of the number of occlusions per frame in each camera view. The higher is the number of occlusions between the detected objects, the higher is the average number of multiple links per object in the corresponding graph, the higher is the value of $\alpha$.
			We can predict $P$ as a function of $\alpha$, of the number of tracked objects $N$, and of the event duration $T$, as:
			\begin{equation}
				P=Ne^{\alpha T}~.
				\label{eqn:mathcalN}
			\end{equation} 
			For $\alpha=0$, the number of paths is exactly equal to the number of real objects, hence $\alpha=0$ corresponds to the ideal case of zero occlusions.
			Eq.~\ref{eqn:mathcalN} is confirmed by tests on synthetic data. In Fig.~\ref{fig5}($a$), the values of $P$ as a function of $T$ are plotted for a synthetic dataset of flocking birds ($N=1024$, see Section~\ref{sec:synthdata} for details). For each $T$, we propagate only the correct temporal links and we measure $P(T)$ for several intervals lasting $T$ frames. The mean value is plotted against $T$, and a linear fit is performed to retrieve the value of $\alpha$. Typically, our experimental data (birds and insects) are characterized by $\alpha\in[0.001,~0.2]$.
		\subsection{Full computational complexity of the problem}
		    The number of paths $P$ is not by itself  the computational bottleneck of the algorithm. What really matters is how many triplets built out of these $P$ paths have a nonzero probability to be stereoscopically connected to each other, because this is what actually enters the global optimization problem. We call $H$ the number of possible stereoscopic matches between the 2D paths across cameras. In the best case scenario, \textit{i.e.} when there are no stereoscopic ambiguities, each one of the 2D paths belongs to only one trajectory (\textit{i.e.}, $\gamma\in\Gamma$), and $H=P$. In the worst case scenario, each one of the 2D paths has at least one stereoscopic link with every other path in the image spaces of the other cameras and, for a three-camera system, $H=P^3$. We can therefore express $H$ as a function of $P$ in the following way, 
			\begin{equation}
				H=P^\beta=\left(Ne^{\alpha T}\right)^\beta~,
				\label{eqn:mathcalH}
			\end{equation}
			where the parameter $\beta\in[1,~3]$ gives the measure of the degree of hybridization between 2D paths on different cameras, which is in turn a function of the optical density of the objects and of the number of occlusions. The longer the temporal acquisition, the higher the probability that one 2D path intersects another one; this is effect is large if there is high diffusion of the real 3D objects in the center of mass reference frame of the group. Therefore, we expect the exponent $\beta$ to grow  with the time duration of the event. This growth of $\beta(T)$ is indeed what we find (see Fig.~\ref{fig5}($b$)); although the saturation limit $\beta\sim2$ is below the upper bound $\beta=3$, the growth of $\beta$ with time means that the exponential explosion of the computational complexity $H$ is rather severe.
		
			In Fig.~\ref{fig5}($c$) we report the computational complexity $H$ as a function of the number of frames $T$ for the same synthetic dataset (yellow diamonds). The exponential growth of $H$ clearly shows that a multi-path branching algorithm by itself cannot solve the matching problem for long time intervals, however optimized it is and however large the memory resources are. Indeed, the very existence of optical occlusions, which is the reason to bifurcate paths in the first place, makes it impossible to reduce significantly the values of $\alpha$ and $\beta$.
		\subsection{Reducing the complexity via recursive divide and conquer}
			The modification through which we are able to drastically decrease the computational complexity of the problem is a recursive divide and conquer strategy.
			
		    The number of 2D paths created in each interval at the first iteration is $P_1=Ne^{\alpha\tau_1}$. The number of possible matches between these paths at the first iteration is $H_1=\left(Ne^{\alpha\tau_1}\right)^{\beta_1}$, where $\tau_1<T$ and $\beta_1=\beta(\tau_1)<\beta(T)$. At the end of the first iteration, the algorithm chooses which 2D paths are kept in memory, and discards the other ones. The number of paths passed at the following iteration is of the same order of $N$ ($N_1\simeq N$ meta-objects are created in each interval). Therefore,
			\begin{equation}
				P_2=N e^{\alpha\tau_2}~~\mathrm{and}~~H_2=\left(N e^{\alpha\tau_2}\right)^{\beta_2}~,
				\label{eqn:mathcalN@iter2}
			\end{equation}
			where $\tau_2<\tau_1<T$ and $\beta_1<\beta_2<\beta(T)$. At the $n$-th iteration,
			\begin{equation}
				P_n=N e^{\alpha\tau_n}~~\mathrm{and}~~H_n=\left(N e^{\alpha\tau_n}\right)^{\beta_n}~,
				\label{eqn:mathcalN@itern}
			\end{equation}
			where $\tau_n<\tau_{n-1}<\cdots<\tau_1<\tau$ and $\beta_1<\beta_2<\cdots<\beta_n\leq\beta(T)$. The crucial point is that $\tau_n\beta_n\ll T\beta(T)$, because we can decide and tune $\tau_n$ to tame the exponential explosion of the computational complexity. Moreover, such strategy allows us balancing the increase of $\beta_i$ from iteration to iteration with a decrease of $\tau_i$. As a result, we can handle a very large number of objects $N$ for an arbitrarily long interval of time, $T$, regardless of the intrinsic complexity of the problem (expressed in terms of $\alpha$ and $\beta$).
			
			In Figure~\ref{fig5}($d$) we plot the computational time versus the problem complexity $H$, with and without the recursive divide and conquer strategy. Thanks to the recursive approach, the computational time is reduced by several orders of magnitude for large values of $H$. Note that, for small values of $H$, the non-recursive approach performs better and should be the preferred choice for small datasets.
			In Figure~\ref{fig5}($e$) we report a quality indicator (MOTA, see next section for its definition) for several runs on synthetic datasets with different complexities $H$, and comparing the results obtained with the global optimization with and without recursive scheme. The plot reveals that the two approaches perform similarly in terms of tracking accuracy.
	\section{Validation with synthetic data}\label{sec:synthdata}
		We validate our algorithm making use of synthetic datasets.
		\\\\
		\textbf{Synthetic data.} We simulate 3D trajectories of flocking birds by adopting a model of self-propelled particles~\cite{bialek2014social}. We use the positions projected in 2D planes directly, rather than generating realistic renderings from them. We do this since we are not interested in testing the performance of the segmentation routine and since it remains hard to predictably simulate the interaction of camera noise and very small objects. Instead, we simulate the errors of the segmentation routine adding white noise directly to the 2D coordinates of the projected objects, in terms of pixel displacements. We also simulate the formation of optical occlusions.
		For the details concerning the generation of the synthetic data, the reader is addressed to Appendix~E. Note though that our simulation still preserves the correspondences between 3D trajectories, the set of 2D projected paths, and the perturbed paths, which can all be used as ground-truth data.
		\\\\
		\textbf{Quality parameters.} Let $G$ be the ground-truth set of trajectories and let $N_G$ be the number of trajectories in $G$. The noisy 2D positions of the ground-truth trajectories are fed to our tracking algorithm, which outputs the set of trajectories $O$. The two sets $G$ and $O$ are compared, and the quality of the output is evaluated in terms of the following parameters:
		\begin{description}
			\item[\textit{MOTA}]: Multiple Object Tracking Accuracy~\cite{keni2008evaluating}, \textit{i.e.} the ratio of the number of correctly reconstructed 3D positions over the total number of 3D positions;
			\item[$G_{90}$]: the ratio of the number of ground-truth trajectories correctly reconstructed for at least $90\%$ frames over the entire event, over $N_G$. For example, given an event lasting $100$ frames, $G_{90}$ represents the percentage of ground-truth trajectories correctly reconstructed for $90$ frames or more).
		\end{description}
		In the best case scenario -- \textit{i.e.} all the ground-truth trajectories are correctly reconstructed -- \textit{MOTA}$=1$ and $G_{90}=1$.
		%
		\begin{table}[!t]
			\renewcommand{\arraystretch}{1.3}
			\caption{\small Summary of the synthetic datasets used to validate the new tracking software. For each dataset, we report its duration expressed in frames, the number of objects $N_G$ of the ground-truth set $G$, the number of output trajectories $N_O$, the value of the parameter $\xi$, and the values of the quality parameters \textit{MOTA} and $G_{90}$.}
			\label{tab:synth}
			\centering
			\begin{tabular}{ccccccccccc}
					\toprule
					Synthetic & Duration & $N_G$ & $N_O$ & $\xi$ & \textit{MOTA} & $G_{90}$ \\
					dataset   & (frames) &       &       &       &               &          \\
					\midrule
					$S1$ &   $125$ &  $256$ &  $256$ & $0.19$ & $0.999$  & $1$     \\
					$S2$ &   $125$ &  $512$ &  $512$ & $0.24$ & $0.9989$ & $0.998$ \\
					$S3$ &   $125$ & $1024$ & $1024$ & $0.27$ & $0.9970$ & $0.987$ \\
					$S4$ &   $250$ &  $256$ &  $257$ & $0.19$ & $0.9975$ & $0.996$ \\
					$S5$ &   $250$ &  $512$ &  $513$ & $0.24$ & $0.9958$ & $0.988$ \\
					$S6$ &   $250$ & $1024$ & $1030$ & $0.27$ & $0.9931$ & $0.967$ \\
					$S7$ &   $500$ &  $256$ &  $257$ & $0.19$ & $0.9989$ & $0.988$ \\
					$S8$ &   $500$ &  $512$ &  $517$ & $0.24$ & $0.9944$ & $0.961$ \\
					$S9$ &   $500$ & $1024$ & $1033$ & $0.27$ & $0.9895$ & $0.928$ \\
					$S10$ & $1000$ &  $256$ &  $257$ & $0.19$ & $0.9991$ & $0.984$ \\
					$S11$ & $1000$ &  $512$ &  $526$ & $0.24$ & $0.9873$ & $0.902$ \\
					$S12$ & $1000$ & $1024$ & $1060$ & $0.27$ & $0.9784$ & $0.869$ \\
					\bottomrule
				\end{tabular}
		\end{table}\linespread{1}
		\\\\
		\textbf{Results on synthetic datasets.} Results for several synthetic datasets are shown in Table~\ref{tab:synth}.
		The quality parameter \textit{MOTA} is always greater than $0.97$, and greater than $0.99$ for 9 datasets over 12. The percentage of correctly reconstructed trajectories is greater than $0.786$. This percentage grows rapidly, as soon as we consider the trajectories which are reconstructed correctly for more than the $90\%$ of the total duration: $G_{90}\geq0.869$.
	\section{Tests on experimental field data}\label{sec:expdata}
		\begin{table*}[!t]
			\renewcommand{\arraystretch}{1.3}
			\caption{\small Summary of the field events analyzed with the new tracking software. For each event, we indicate the object type, the estimated number of objects, the duration (in frames and seconds), the acquisition frame-rate, and the percentage of reconstructed trajectories whose length is greater than $90\%$ of the acquisition duration.}
			\label{tab:exp}
			\centering
			\begin{tabular}{ccccccc}
					\toprule
					Experimental & ~Object~ & ~~Estimated~~ & ~~~Duration~~~ & ~Frame-rate~~ & ~$\%$ of trajectories~ \\
					  dataset    &   Type   & $\#$ Objects  & (frames~$|$~s) &     (Hz)      &  with Length $>90\%$   \\
					\midrule
					$E1$   & birds   &  $179$ &   $440~|~5.50$  &  $80$ & $87.0\%$\\ 
					$E2$   & birds   &  $551$ &   $360~|~4.50$  &  $80$ & $90.2\%$\\ 
					$E3$   & birds   &  $365$ &   $128~|~1.60$  &  $80$ & $78.6\%$\\ 
					$E4$   & birds   &  $120$ &   $310~|~1.82$  & $170$ & $99.2\%$\\ 
					$E5$   & birds   &   $50$ &  $1000~|~5.88~$ & $170$ & $98.0\%$\\ 
					$E6$   & birds   &  $482$ &   $761~|~4.48$  & $170$ & $84.3\%$\\ 
					$E7$   & birds   &  $117$ &   $500~|~2.94$  & $170$ & $88.7\%$\\ 
					$E8$   & birds   &  $110$ &   $661~|~3.89$  & $170$ & $97.2\%$\\ 
					$E9$   & birds   &  $381$ &   $960~|~5.65$  & $170$ & $72.3\%$\\ 
					$E10$  & birds   &  $168$ &   $300~|~1.76$  & $170$ & $81.2\%$\\ 
					$E11$  & birds   & $1270$ &   $300~|~1.76$  & $170$ & $87.6\%$\\ 
					$E12$  & birds   &   $60$ &   $609~|~3.58$  & $170$ & $89.8\%$\\ 
					$E13$  & insects &   $37$ & $2000~|~11.76$  & $170$ & $97.1\%$\\ 
					$E14$  & insects &  $332$ &  $465~|~2.73$   & $170$ & $80.2\%$\\ 
					$E15$  & insects &  $115$ & $1000~|~5.88~$  & $170$ & $85.6\%$\\ 
					$E16$  & insects &  $147$ & $1000~|~5.88~$  & $170$ & $84.6\%$\\ 
					$E17$  & insects &  $210$ &  $500~|~2.94$   & $170$ & $82.7\%$\\ 
					$E18$  & insects &  $124$ & $1024~|~6.02~$  & $170$ & $84.0\%$\\ 
					$E19$  & insects &  $633$ &  $250~|~1.47$   & $170$ & $82.3\%$\\ 
					\bottomrule
				\end{tabular}
		\end{table*}\linespread{1}
		We also tested our algorithm using our experimental data of flocking birds and swarming insects acquired on the field, as well as using public benchmark datasets.
		
		Testing the algorithm with our data, we analyzed $12$ events of starling flocks~\cite{attanasi2014information}, and $7$ events of swarming midges~\cite{attanasi2014collective,attanasi2014prl}, as summarized in Table~\ref{tab:exp}.
		Fig.~\ref{fig1} shows the reconstructed trajectories for four events, the bird flocks labelled $E3$ and $E6$, and the midge swarms labelled $E14$ and $E15$ -- see Table~\ref{tab:exp}. The original video sequences of these four experimental events (in slow-motion, $0.15\times$~slower than the original speed), together with the reconstructed trajectories, are included as Supplemental Material.
		Clearly, ground-truth trajectories are not available in the case of experimental data, and -- due to the size of our datasets -- manual inspection is not feasible, except for a limited number of ambiguous cases. The quality of the reconstructed trajectories is assessed in this case only in terms of trajectory fragmentation. In Table~\ref{tab:exp}, we report the percentage of trajectories longer than the $90\%$ of the duration of the acquisition. The majority of the reconstructed trajectories are of full-length, and trajectory fragmentation is negligible. Such high-quality data have been used to perform the analysis presented in~\cite{attanasi2014information} and~\cite{attanasi2014collective,attanasi2014prl}.
		
		\begin{table}[!b]
            \renewcommand{\arraystretch}{1.3}
            \caption{\small Comparison of the quality of the output trajectories retrieved using GReTA and the ones retrieved using the algorithms MHT and SDD-MHT, as published by Z.~Wu~\textit{et~al.}~\cite{wu2014thermal} (see Table~IV therein) on the datasets labeled \textit{Davis-08 sparse} and \textit{Davis-08 dense}.}
            \label{tab:wu}
            \centering
            \begin{tabular}{clccccc}
                    \toprule
                    Dataset  		 & Algorithm &   MT      &    ML    &   FM   &  IDS   &       \textbf{MOTA}         \\
									 &           & ($\%$)    &  ($\%$)  & ($\#$) & ($\#$) &      ($\%$)        \\
                    \midrule
                      & MHT       & $96.6$  & ~$0$   & $105$  & ~$97$  & ~$\mathbf{64.1}$ \\
                    \textit{sparse}  & SDD-MHT   & $95.2$  & ~$0$   & $145$  & $126$  & ~$\mathbf{78.9}$ \\
                      & GReTA       & $83.1$  & $3.4$  & $188$  & ~~$9$  & ~$\mathbf{82.4}$ \\
                    \\
                      & MHT       & $71.9$  & $2.5$  & $274$  & $355$  &  $\mathbf{-32.0}$~ \\
                    \textit{dense~}  & SDD-MHT   & $61.1$  & $3.0$  & $454$  & $444$  & ~$\mathbf{44.9}$ \\
                      & GReTA       & $78.8$  & $2.9$  & $335$  & ~~$8$  & ~$\mathbf{80.3}$ \\
                    \bottomrule 
                \end{tabular}
        \end{table}\linespread{1}
		To the best of our knowledge, the only public benchmark datasets for 3D-tracking of animal groups are the thermal infrared videos (the raw image sequences with the corresponding sets of ground-truth trajectories) published by Z. Wu and coworkers~\cite{wu2014thermal}. We tested our tracking algorithm on the two datasets \textit{Davis08-sparse} and \textit{Davis08-dense}, and we evaluated the output trajectories using the quality parameters defined in~\cite{wu2014thermal}: the numbers of Mostly Tracked ($MT\geq80\%$) trajectories, Mostly Lost ($ML\leq20\%$) trajectories, track fragmentations (FM) and Identity Switches (IDS), as well as the MOTA. In Table~\ref{tab:wu}, we report the quality of our output trajectories compared to the results on the same datasets published by Z. Wu~\textit{et~al.}~\cite{wu2014thermal}.

		Our results exhibit better values of MOTA and IDS on both datasets, \textit{dense} and \textit{sparse}, revealing that the trajectories are characterized by low values of false positives, identity switches and mismatches. In terms of MT, ML and FM, GReTA performs slightly worse on the \textit{sparse} dataset than MHT and SDD-MHT; on the other hand, when applied to the \textit{dense} dataset, its performance is comparable to the one of the other methods.
		This implies that MHT and SDD-MHT output a larger percentage of complete trajectories, which nevertheless are characterized by more identity switches and false positives -- as revealed by the lower values of MOTA and higher values of IDS. We were not surprised by this, as our tracking algorithm is intentionally designed to discard false positives, preferring short and correct trajectories to long but incorrect ones.
		
		We believe that this benchmark proves the performance advantages, as well as the flexibility of GReTA to process very diverse experimental data.

	\section{Comparison with prior work}\label{sec:comp}
\begin{figure}[!b]
			\centering
			\includegraphics[width=0.90\columnwidth]{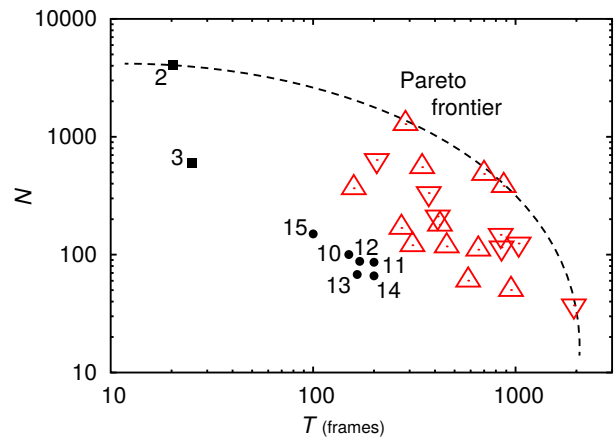}
			\caption{\small Comparison of test-cases used for several tracking algorithms, quantified in terms of temporal duration $T$ and estimated number of tracked objects $N$. The largest datasets processed by different tracking algorithms define the Pareto frontier in the two-dimensional space $\lbrace T,~N\rbrace$. The points are classified according to the field of investigation for which the respective algorithms have been developed: fluid dynamics experiments ({\tiny$\blacksquare$}), biological experiments ($\bullet$), and the experimental data processed using GReTA and presented in this paper and listed in Table~\ref{tab:exp}
			(\textcolor{red}{$\triangle$}~and~\textcolor{red}{$\bigtriangledown$} for birds and insects, respectively). The numbers next to the symbols correspond to the references to the papers from which $T$ and $N$ have been estimated -- see Bibliography.}
			\label{fig6}
		\end{figure}

		In order to situate our algorithm in the 3D tracking landscape, an estimate (when explicit information was not published) of the number of tracked objects $N$ and of the temporal duration $T$ (the average trajectory length is used in case of objects entering and leaving the field-of-view) is shown in Fig.~\ref{fig6} for a number of 3D tracking results published in the literature. The points scattered on the plot have been classified according to the field of investigation for which the respective algorithms have been developed: fluid dynamics experiments ({\tiny$\blacksquare$}), biological experiments ($\bullet$), and the experimental data presented in this paper and listed in Table~\ref{tab:exp} (\textcolor{red}{$\triangle$}~and~\textcolor{red}{$\bigtriangledown$} for birds and insects, respectively). The numbers next to the symbols correspond to the references to the papers from which $T$ and $N$ have been estimated -- see Bibliography.
		Given that $N$ and $T$ are both valid -- but qualitatively different -- criteria of evaluation, to compare different methods according to $N$ and $T$ we can use a multi-objective optimization approach, the simplest of which is defining the Pareto frontier~\cite{pareto1964book,messac2003smo} in the $\lbrace T,~N\rbrace$ plane.
		We sketched with a dashed line the Pareto frontier for the plotted data-points in the 2D space of $N$ and $T$. The best tracking performance is given by the points closest to the frontier.
		
		The plot clearly shows that fluid dynamics tracking algorithms have been optimized to track large number of tracer particles for short times, and that previous tracking approaches restricted biological experiments to relatively small numbers of animals, even though for longer times. The data processed using GReTA mark the Pareto frontier, assessing the important step forward in terms of performance of the algorithm we proposed. This proved to be suitable for tracking large groups of objects for considerably long durations, without suffering from frequent optical occlusions.
	\section{Conclusions}\label{sec:conclusions}
		We presented a novel Global and Recursive Tracking Algorithm (GReTA) in three dimensions able to reconstruct uninterrupted trajectories for large numbers of objects and long time intervals, even with frequent optical occlusions. This recursive divide and conquer algorithm is based on the idea of global optimization of the solution -- global in space as well as in time. The applicability of a global optimization is limited by the computational complexity, which grows exponentially fast with the time duration of the sequence.
		
		Here we achieve a dramatic reduction of the computational complexity by making use of a recursive divide and conquer strategy, which allows to first optimize the matches globally over shorter temporal intervals, and then iterate to cover the entire temporal sequence. In this way, the computational complexity is drastically reduced while preserving the global scope, permitting to track very large datasets (large in terms of number of objects and of duration of the video acquisition). We further proposed several adaptations making the algorithm robust against wrong or missing links.

		We implemented the algorithm; we validated it making use of synthetic data with available ground-truth information; we tested it on new experimental field data of flocking birds and swarming insects; we compared its performance using public benchmark datasets. We showed that the algorithm is capable of reconstructing 3D-trajectories with negligible fragmentation, and that the quality of the trajectories is not affected by the recursive divide and conquer strategy. To the best of our knowledge, the results based on synthetic data and on the public datasets proved the superior performance of the proposed tracking approach compared to other existing methods.

		We processed bird flock data, insect swarm data, and bats data, despite these systems being very different from each other: insects in a swarm fly in a very jerky manner and occlude frequently in the images, but for very short times; the flight of birds in a flock is highly coordinated, so that occlusions are typically very long-lasting, and can involve several birds at the time; bats exiting a cave continuously enter and leave the field-of-view. Because of this flexibility, we believe that the GReTA approach can be successfully applied to process the most diverse experimental data.
\ifCLASSOPTIONcompsoc
	\section*{Acknowledgments}
\else
	\section*{Acknowledgment}
\fi
		This work was supported by grants IIT--{Seed Artswarm}, ERC--{StG n.257126}, and AFOSR--{FA95501010250} (through the University of Maryland). F.~Pellacini was partially supported by Intel. We acknowledge the advice of Carlo Lucibello on multi-objective optimization.
\end{document}